\documentclass[aps,prb,twocolumn,floatfix]{revtex4}
\usepackage[final]{graphicx}
\usepackage{bm}
\usepackage{float}
\usepackage{afterpage}

\begin{document}

\title{Quantum pumping induced by disorder in one dimension}

\author{Jihong Qin}
\affiliation{Department of Physics, University of Science and
Technology Beijing, Beijing 100083, China}
\author{Huaiming Guo}
\email[Corresponding author. {\it E-mail address}: ]{hmguo@buaa.edu.cn}
\affiliation{Department of Physics, Beihang University, Beijing 100191, China}

\begin{abstract}
The topological property in one dimension is protected by symmetry. Based on a concrete model, we study the effect of disorder preserving or breaking the symmetry and show the nature of symmetry protecting in the one dimension topological phase. A stable quantum pumping can be constructed within the topological model. It is shown that an integer charge is pumped across a periodic chain in a cyclic process. Furthermore we find that not only the quantum pumping is stable to on-site disorder, but also can be induced by it. These results may be realized experimentally using quasicrystals.
\end{abstract}

\pacs{
            }
\maketitle


\section{Introduction}

The discovery of topological insulators (TIs) have aroused the interests in the study of topological phase of matter \cite{rev1,rev2,rev3,rev4}. Many interesting physical properties are predicted and different experimental setups are devoted to study them. Besides the experiments on real materials, the interest also spreads to the fields of ultracold atoms and photonic crystals, in which the model Hamiltonian can be directly engineered in a highly controllable and tunable way. Remarkably great progress has been achieved in these fields, such as: the realizations of synthetic gauge field \cite{sm1,sm2,sm3} and the spin-orbit coupling \cite{sp1,sp2,sp3} in ultracold atoms, adiabatic pumping in quasicrystals, photonic graphene, photonic Floquent TIs, et al \cite{p1,g1,f1}.

An exhaustive classification shows that in every spatial dimension there exist precisely five distinct classes of topological insulators or superconductors. Compared to the case in two (2D) or three dimensions (3D), the physics in one dimension (1D) is relatively easier to understand. The 1D topological property was studied as early as 1979, when  Su et al. suggest a famous model to study the solitons in polyacetylene , known as Su-Schrieffer-Heeger (SSH) model \cite{ssh}. Recently a lot of works refocus on 1D.
Many phenomena predicted in 2D or 3D are also suggested in 1D, such as the topological phase transition, topological Mott insulator, fractional topological phase, et al \cite{tp1,tp2,tp3,guo1,guo2,qp1}. Since it is already possible to realize some of the underlying models in the field of ultracold atoms, these phases can be studied experimentally \cite{exp1,exp2,exp3}.
Besides the studies on fermions, the studies are also extended to bosons \cite{guo3,bs1,bs2,bs3}, which are more available in the kind of experiments. A recent experiment using light propagating in photonic waveguides has simulated a kind of 1D topological model. Some important topological properties are exhibited using laser. Another 1D topic attracts many studies is about Majorana fermions, which may have important applications in topological quantum computing \cite{mj}.

Due to the intense interests in 1D topological phase, in this paper,  we systematically study a 1D topological model and the quantum pumping based on it. Specially we find that not only the quantum pumping is stable to on-site disorder, but also can be induced by it. This result is very possible to be realized in the present experimental setup. The paper is organized as follows. In Sec.II, a 1D topological model reduced from the Benevig-Huges-Zhang (BHZ) model is introduced and the effect of the disorder on the topological phase is studied. In Sec. III, we show that based on the 1D topological model a stable quantum pumping can be constructed and demonstrate it from several aspects. In Sec. IV, the quantum pumping induced by the disorder is studied. Finally we conclude and discuss the possible experimental realization of these results in Sec.V.

\section{The 1D topological phase}

Our starting point is the 1D non-interacting tight-binding model,\cite%
{guo1,guo2,guo3}
\begin{eqnarray}
\label{eq1}
H_{0} &=&\sum_{i}(M+2B)\Psi _{i}^{\dagger }\sigma _{z}\Psi _{i}-\sum_{i,\hat{%
x}}B\Psi _{i}^{\dagger }\sigma _{z}\Psi _{i+\hat{x}} \nonumber  \\
&-&\sum_{i,\hat{x}}sgn(\hat{x})IA\Psi _{i}^{\dagger }\sigma _{x}\Psi _{i+%
\hat{x}} \label{eq1}
\end{eqnarray}%
where $\sigma _{x}$, $\sigma _{z}$ are Pauli matrices; $I$ the imaginary unit and $\Psi
_{i}=(c_{i},d_{i })^{T}$ with $c_{i }$($%
d_{i }$) electron annihilating operator at the site $\mathbf{r}%
_{i} $. The first term is the on-site potential, which has different signs for the s-orbit and d-orbit.  The second term is the hopping amplitudes among the s-orbits or d-orbits, which are also differed by a sign. The third term is the hopping amplitudes between the s-orbit and d-orbit electrons, which is due to the spin-orbit coupling. In the following of the paper we take $B$ positive and set $A=1$ as
the energy scale.

In momentum space Eq.(\ref{eq1}) becomes $H_{0}=\sum_{k}\Psi
_{k}^{\dagger }\mathcal{H}(k)\Psi _{k}$ with $\Psi _{k}=(c_{k
},d_{k })^{T}$ the Fourier partner of $\Psi _{i}$ and
\[
\mathcal{H}(k)=[M+2B-2Bcos(k)]\sigma _{z}+2Asin(k)\sigma _{x}.
\]%
The spectrum of $\mathcal{H}(k)$ consists of two bands,
\[
E_{k}^{(1,2)}=\pm \sqrt{\lbrack M+2B-2Bcos(k)]^{2}+[2Asin(k)]^{2}}.
\]%
At half-filling, the
system can be a non-trivial insulator with edge
modes for $-4B<M<0$  and a trivial insulator for $M>0$ or $M<-4B$.
The topological property of the system can be understood in terms of Berry
phase in $k$ space, which is $\gamma =\oint \mathcal{A}(k)dk$ with the Berry
connection $\mathcal{A}(k)=i\langle u_{k}|\frac{d}{dk}|u_{k}\rangle $ and $%
|u_{k}\rangle $ the occupied Bloch state\cite{berry}.
In the Hamiltonian Eq.(\ref{eq1}) only two of three Pauli matrices are used and the left one $\sigma_y$ automatically becomes the symmetry protecting the topological phase: ${\cal H}(k)=-\sigma_y {\cal H}(k) \sigma_y$. So the Berry phase $\gamma $ mod
$2\pi $ can have two values: $\pi$ for the topological phase and $0$ for the trivial phase.

\begin{figure}[htbp]
\centering
\includegraphics[width=8.5cm]{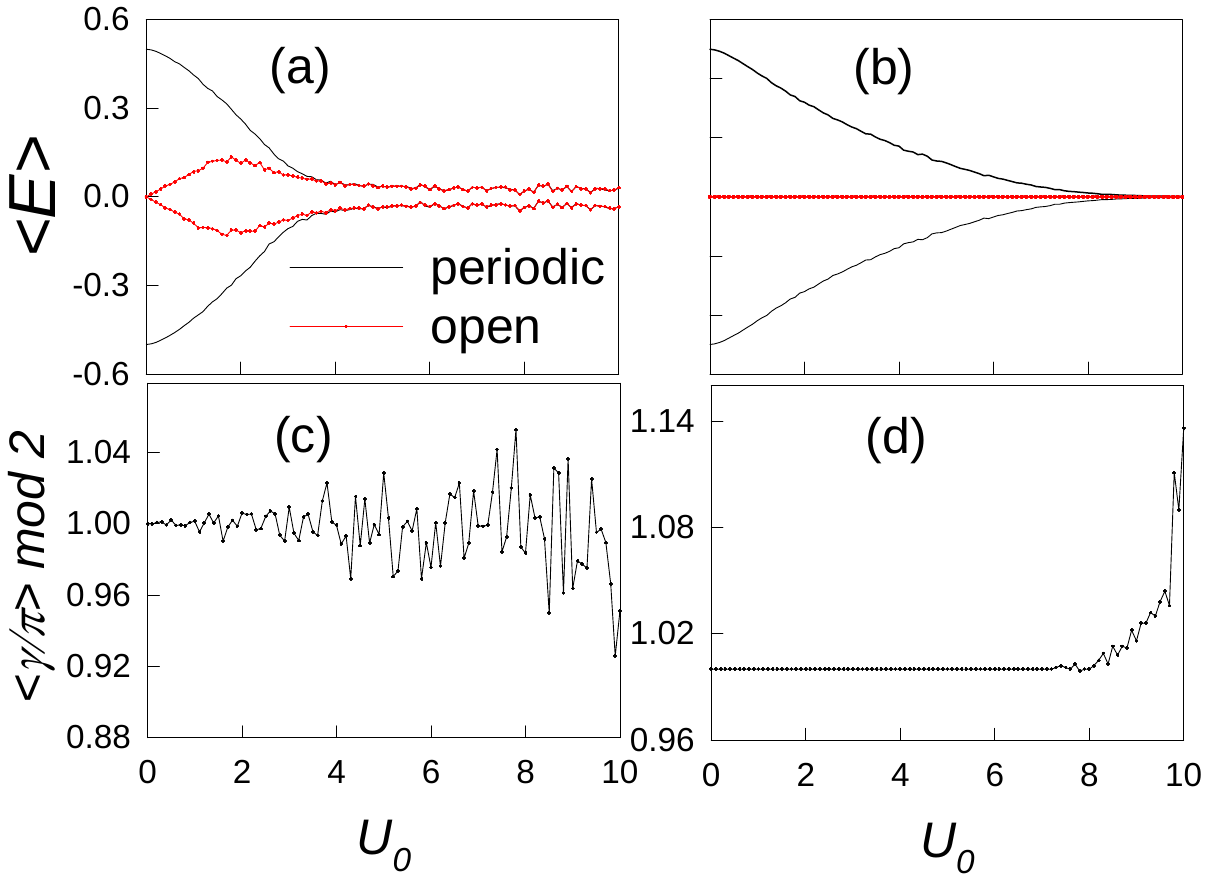}
\caption{(Color on line) The average of the $N$ and $N+1$ eigenenergies with the disorder in: (a) $\sigma_0$ channel; (b) $\sigma_z$ channel. (c) and (d) are the corresponding Berry phase of (a) and (b), respectively. The parameters are: $B=1$, $M=-0.5$ and the length of the chain $N=100$. The result is averaged over $1000$ different configurations of the disorder. }\label{fig1}
\end{figure}

When terms violating the above symmetry are added to Eq.(\ref{eq1}), the topological phase will be broken, such as terms in the channels of $\sigma_0, \sigma_y$ with $\sigma_o$ the $2\times 2$ unit matrix and $\sigma_y$ the Pauli matrix. Since the on-site disorder with $\sigma_0 -$ channel usually exists in real system, the 1D topological phase is fragile. Next we add the on-site disorder, which writes as,
\begin{eqnarray}
H_{dis}=\sum_{i} U_i (c^{\dagger}_{i} c_{i}+d^{\dagger}_{i} d_{i}),
\end{eqnarray}
with $U_i$ uniformly distributed in $(-\frac{U_0}{2},\frac{U_0}{2})$.
As shown in Fig.\ref{fig1} (a) and (c), in the topological phase of Eq.(\ref{eq1}), once the on-site disorder is added, the two degenerate zero modes on an open chain will be gapped, implying the topological property is broken. Also the Berry phase without symmetry protecting can take arbitrary values. To make a comparison, we calculate the case with $\sigma_z -$ channel disorder, which is shown in Fig.\ref{fig1} (b) and (d). This kind of disorder preserves the symmetry of the clean system. The results show that to very large strength of the disorder, the zero modes persist and the Berry phase $\gamma $ mod $2\pi $ is quantized to $\pi$.  The above results show that the 1D topological phase only is robust when the protecting symmetry is preserved.

\begin{figure}[htbp]
\centering
\includegraphics[width=8cm]{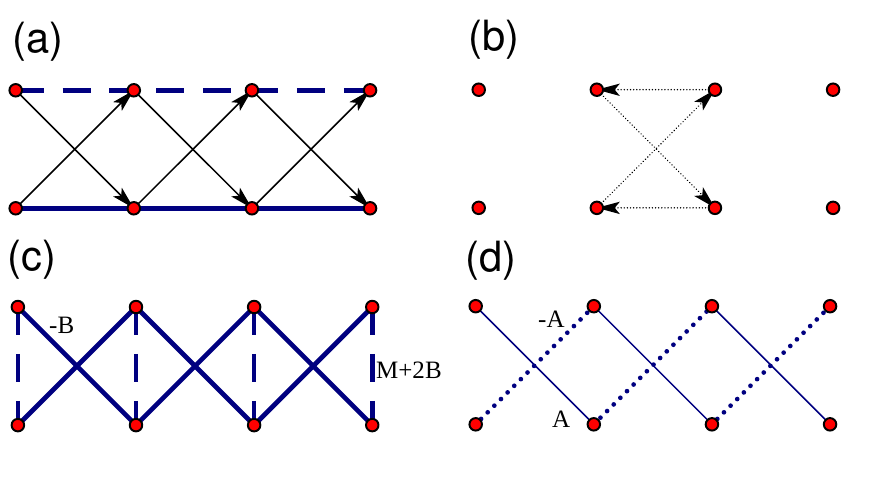}
\caption{(Color online) (a) Schematic representation of the Hamiltonian Eq.(\ref{eq1}), in which the dashed line represents a hopping with a negative sign; the line with an arrow represents an imaginary hopping and the hopping along the arrow is positive. (b) the eigenstate of Eq.(\ref{eq1}) under the parameters: $B=1,M=-2B$. The arrow represents the direction of the electrons' hopping. The schematic representation of the Hamiltonian Eq.(\ref{eq6}) in real space: (c) the $\sigma_x$ term, (d) the $\sigma_y$ term. The $c-$ and $d-$ orbit on the same site is represented by the lower and upper red dots, respectively.}\label{fig2}
\end{figure}

To further understand the above result, we present a special case when $B=1$ and $M=-2B$.  Under this set of parameters, the bands are flat with the eigenvalues $-2$ and $2$. The hoppings of the electrons on the chain are schematically shown in Fig.\ref{fig2} (a). It is found that in the presence of imaginary hopping the path along one direction (determined by the sign of the imaginary hopping) is energetically favored. Thus for the special case, an electron hopping along a close loop between two adjacent sites is an eigenstate of the system. Since the electron is localized between the two adjacent sites, its Hamiltonian is expressed as:
\begin{eqnarray*}
h_{i,i+1}=\left(
      \begin{array}{cccc}
   0 & 0 & -B & -I A \\
  0 & 0 & -I A & B \\
  -B & I A & 0 & 0 \\
  I A &  B & 0 & 0
      \end{array}
    \right).
\end{eqnarray*}
Two of its eigenstates: $|\psi_1\rangle=\{-0.5I,0.5,-0.5I,-0.5\}^{T}$ with the eigenvalue $-2$ and $|\psi_2 \rangle=\{0.5I,-0.5,-0.5I,-0.5\}^{T}$ with the eigenvalue $2$, are also the eigenstates of the whole system (the other components in the enlarged basis are all zeros). On an open chain with $N$ sites, $N-1$ such states with the eigenvalue $-2$ firstly be filled into the system, resulting a distribution with one particle on each inner site and half on each edge site. Before filling $N-1$ above states with the eigenvalue $2$, there appear two zero modes which exactly localized at the edges. Their wavefunctions are: $|\psi_1(0)\rangle=\frac{1}{\sqrt{2}}\{i,1\}^{T}$ (the components on site $i\neq 1$ are zeros) and $|\psi_2(0)\rangle=\frac{1}{\sqrt{2}}\{-i,1\}^{T}$ (the components on site $i\neq N$ are zeros). In the presence of $\sigma_0 -$ channel disorder, suppose the strength of the random on-site potentials on sites $i=1$ (the first site) and $i=N$ (the end site) are $U_1$ and $U_N$ ($U_1 \neq U_N$), respectively. Though the eigenstates of the clean system are still the eigenstates of the disordered one, the eigenenergies of the two zero modes become $U_1$ and $U_N$, which are no longer degenerate. While for the disorder in $\sigma_z -$ channel, the eigenenergies of the two zero modes are not altered.
When the bands have dispersion, the situation becomes more complex, however the same physics remains. So through the special case, the robustness of 1D symmetry protected topological phase is intuitively understood.
\section{Quantum pumping and its robustness to disorder}
\begin{figure}[htbp]
\centering
\includegraphics[width=9cm]{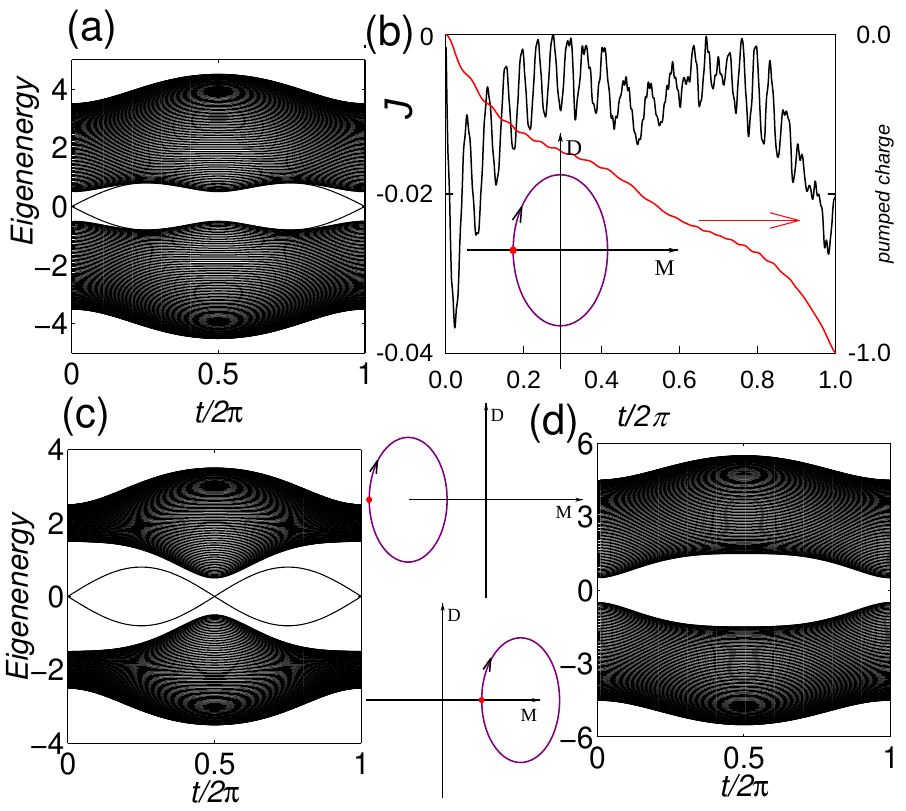}
\caption{(Color on line) The instantaneous energy spectrum on an open chain along the loop: (a) $M_1=0$; (c) $M_1=-1$ and (d) $M_1=1$. (b) the instantaneous local current and the pumped charge of the same process with (a) on a periodic chain at half-filling. The other parameters are: $B=1$,$M_0=-0.5$, $D_0=0.8$ and the length $N=100$.}\label{fig3}
\end{figure}

As shown in the previous section, due to the presence of one free Pauli matrix, the 1D topological phase is usually fragile to the on-site disorder. A natural thought is to include a term in $\sigma_y -$ channel in Eq.(\ref{eq1}), which leads to the quantum pumping, i.e., the cyclic adiabatic evolution of 1D insulator \cite{qp1,bs2,qp2,qp3,qp4}. Next we add an on-site term  $H'=-D \sum_{i}(Ic^{\dagger}_{i}d_{i}-Id^{\dagger}_{i}c_{i})$ to Eq.(\ref{eq1}), which is in the $\sigma_y -$ channel. Then we have a parameters' plane $(M,D)$. If we start from a point $(M_0,0)$ (let $-4B<M_0<0$ when the 1D system is in the topological phase) in the plane and draw a close loop round $(0,0)$, a quantum pumping process is created, as shown in Fig.\ref{fig3}. In the following we take the path $(M,D)=(M_0 \cos 2\pi t/T, D_0 \sin 2\pi t/T)$ with $t$ a variable (it can be viewed as time) and $T$ the period of the cycle. In momentum space,
\begin{eqnarray}\label{eq3}
\mathcal{H'}(k,t)&=&[M+2B-2Bcos(k)]\sigma _{z} \ \\ \nonumber
&+&2Asin(k)\sigma _{x}+D \sigma_y.
\end{eqnarray}

The instantaneous energy spectrum is calculated and is shown in Fig.\ref{fig3}(a). The system is gapped on the periodic chain. As the boundary condition is changed to the open one, there appears gapless states traversing the gap, which is very similar to the energy spectrum of two-dimensional topological phase. However the in-gap states represent the evolution of the edge modes.
During the pumping process, the time evolution of the wavefunction $|\psi (t)\rangle$ can be calculated by step-vise change of $t$ in small increments $\delta t$ \cite{exact},
\begin{eqnarray}\label{eq4}
|\psi (t+\delta t)\rangle \simeq \sum_{l=1}^{2N} e^{-i\epsilon_{l} \delta t} |\phi_l\rangle  \langle \phi_l|\psi(t)\rangle,
\end{eqnarray}
where $|\phi_l\rangle ,\epsilon_l, (l=0...2N)$ are the static eigenfunction and eigenenergy of the total Hamiltonian at $t$. With the wavefunction $|\psi (t)\rangle$, the pumping process can be demonstrated on a chain with periodic (PBC) or open boundary condition (OBC).

For a cyclic process on a periodic chain, the charge pumped at half filling across the insulator is a integer, which can be defined as a topological invariant.
To show the details of the process, we directly calculate the current and the number of particles which have flown through the chain between the starting time and time $t$  is expressed as the integral of the current ${\cal J}_j$ at site $j$: $\Delta n(t)=\int_0^T dt' {\cal J}_j(t')$, where ${\cal J}_j=\langle \psi(t)|\hat{{\cal J}}_j|\psi(t)\rangle$ being the expectation value of the current operator $\hat{{\cal J}}_j$. The operator $\hat{{\cal J}}_j$ can be obtained vis the continuity equation \cite{note1}. For the above process, $\hat{{\cal J}}_j=\hat{{\cal J}^{cc}}_j+\hat{{\cal J}^{dd}}_j+\hat{{\cal J}^{cd}}_j+\hat{{\cal J}^{dc}}_j$ with $\hat{{\cal J}^{cc}}_j=-I (-B) (c_j^{\dagger}c_{j-1}-c_{j-1}^{\dagger}c_{j})$, $\hat{{\cal J}^{dd}}_j=-I (B) (d_j^{\dagger}d_{j-1}-d_{j-1}^{\dagger}d_{j})$, $\hat{{\cal J}^{cd}}_j=-I (I A) (c_j^{\dagger}d_{j-1}+d_{j-1}^{\dagger}c_{j})$ and $\hat{{\cal J}^{dc}}_j=-I (I A) (d_j^{\dagger}c_{j-1}+c_{j-1}^{\dagger}d_{j})$. In Fig.\ref{fig3} (b) the instantaneous local current and the total pumped charge on a system with PBC at half filling is shown and the result shows that one charge is pumped in a cycle.

\begin{figure}[htbp]
\centering
\includegraphics[width=9cm]{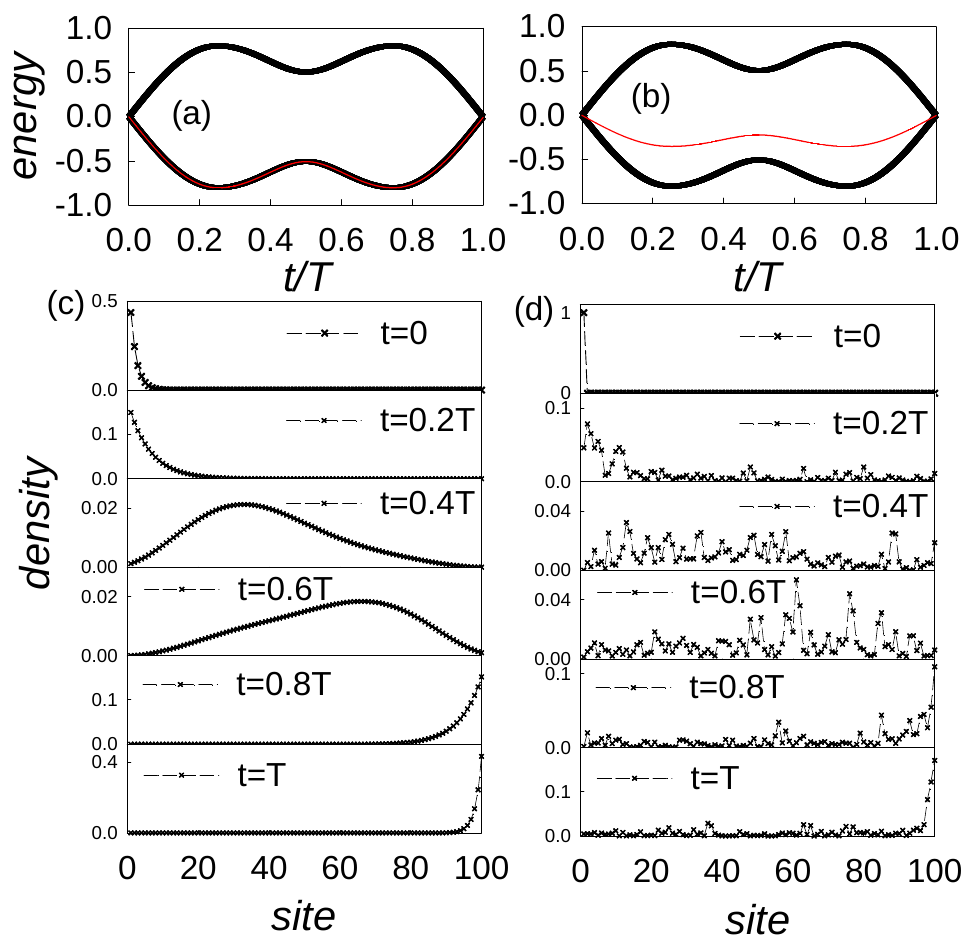}
\caption{(Color on line) (a) and (b): the average energy of the time-evolution wavefunction (the red color one, the black ones are the gapless states in the instantaneous energy spectrum of Fig.\ref{fig3} (a) ). (c) and (d): its distribution. (a) and (c) correspond to the case of starting from an eigenstate near the edge, while (b) and (d) correspond to the case of starting from a state with one particle placed on the left edge site. }\label{fig4}
\end{figure}

\begin{figure}[htbp]
\centering
\includegraphics[width=9cm]{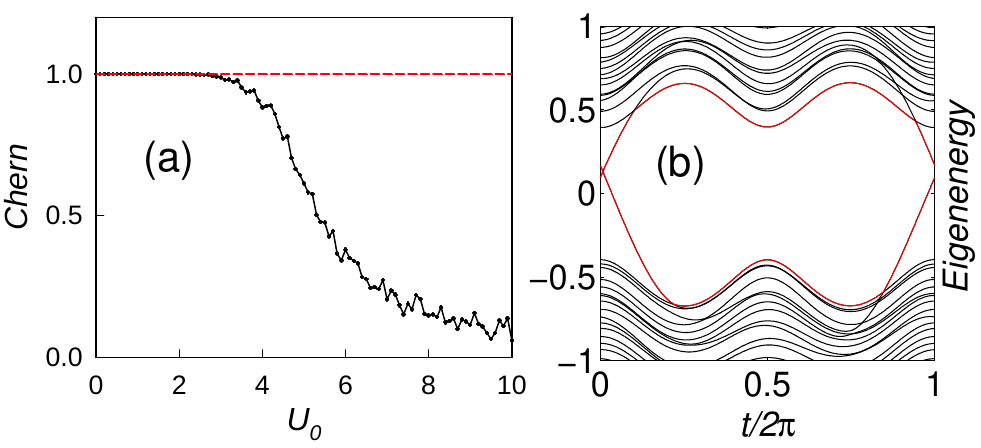}
\caption{(Color on line) (a) The Chern number vs. the strength of the on-site disorder. (b) the low-energy instantaneous energy spectrum with $U_0=1$, where the red lines are the $N-$ and $N+1 -$ eigenenergies. The clean pumping process is the same with that in Fig.\ref{fig3}(a) and the Chern number is averaged over $1000$ different configurations of disorder. }\label{fig5}
\end{figure}

On an open chain the pumping process can be demonstrated through the adiabatic pumping of the edge state. Firstly we start from the $N-th$ eigenstate $|\psi\rangle_N$ (the one at half filling) at $t=0$, which is an edge state distributing mainly near the left edge (Fig.\ref{fig4}(c)). According to Eq.(\ref{eq4}), its time evolution $|\psi (t)\rangle_N$ and the distribution $n_i=\langle\psi (t)|\hat{n_i}|\psi(t)\rangle_N$ can be calculated, where $\hat{n_i}=c_i^{\dagger}c_i+d_i^{\dagger}d_i$ is the operator of the particle number at the $i-th$ site. We show the distribution of the particle at several typical times in Fig.\ref{fig4}(c). In one period, the edge state is pumped from the left edge to the right one. In the process the average energy of the evolving state $\langle E\rangle_t=\langle\psi (t)|\hat{H'}|\psi(t)\rangle_N$ with $\hat{H'}$ the Hamiltonian in real space corresponding to ${\cal H'}(k,t)$ (see Eq.(\ref{eq3})) is calculated. As shown in Fig.\ref{fig4}(a), it is along the gapless state traversing the gap in the energy spectrum shown in Fig.\ref{fig3}(a). So on an open chain the particle placed on one edge can be pumped to the other edge through the quantum pumping, which has been observed in the experiments using quasicrystals \cite{p1}. We also simulate a more realistic process based on our model. As shown in Fig. \ref{fig4}(d), we place a particle on the edge site, which is not the eigenstate of the system. After the pumping of one period, its main part is pumped to the other edge.

For a periodic system, varying the momentum $k$ and the time $t$, we get a manifold of Hamiltonian ${\cal H'}(k,t)$. If in a cycle the system remains gapped, the Chern number can be defined as \cite{ch1,ch2}:
\begin{eqnarray}\label{eq5}
C=\frac{1}{2\pi i}\int_{0}^{2\pi}dk \int_{0}^{T}dt F_{12}(k,t).
\end{eqnarray}
Here the field strength $F_{12}(k,t)=\partial_{k} A_2(k,t)-\partial_{t} A_1(k,t)$ with the Berry connection $A_{1(2)}(k,t)=\langle n(k,t) |\partial _{k (t)}|n(k,t) \rangle$ and $|n(k,t)\rangle$ the normalized wave function.  Using the numerical approach suggested in Ref.(36), the Chern number of the system in Fig.\ref{fig3}(a) is calculated. We find that the Chern number are $1$ at half filling. The Chern number is consistent with the appearance of gapless states and the pumped integer charge, which manifest the nontrivial topology in the system.

Generally the topological property of a quantum pumping is only determined by the condition whether only one of the crossings between the loop and the $M-$ axis is in $(-4B,0)$ and it does not depend on its specific shape, which is a key feature of the topological property. When the above loop is changed, different energy spectrum can be obtained. For example, we move the path along $M-$ axis: $(M,D)=(M_1+M_0 \cos t, D_0 \sin t)$ with $M_1$ the displacement. When $M_1+M_0$ and $M_1-M_0$ are both in $(-4B,0)$, the in-gap states have two crossings (Fig.\ref{fig3}(c) ). When $M_1+M_0$ and $M_1-M_0$ are both bigger than $0$, the in-gap states disappear (Fig.\ref{fig3}(d) ). However these two cases correspond to trivial cases, in which the pumped charge is zero and the Chern number is zero.

Since the Pauli matrices are used up in describing the quantum pumping, the quantum pumping is robust to on-site disorder. We calculate the Chern number in the presence of disorder and the result is shown in Fig.\ref{fig5}. The Chern number averaged on $1000$ different configurations of the disorder is quantized to $1$ to quite large strength of the disorder. The critical strength is related to the size of the gap and it has larger value for the system with bigger gap. Though small disorder doesn't destroy the topological property of the system, it affects the energy spectrum. As shown in Fig.\ref{fig5} (b), the details of the energy spectrum is changed by the disorder. However the in-gap states with one crossing induced by the topological property remains and an integer charge is still pumped in the disordered process.

\section{Quantum pumping induced by disorder}

\begin{figure}[htbp]
\centering
\includegraphics[width=7cm]{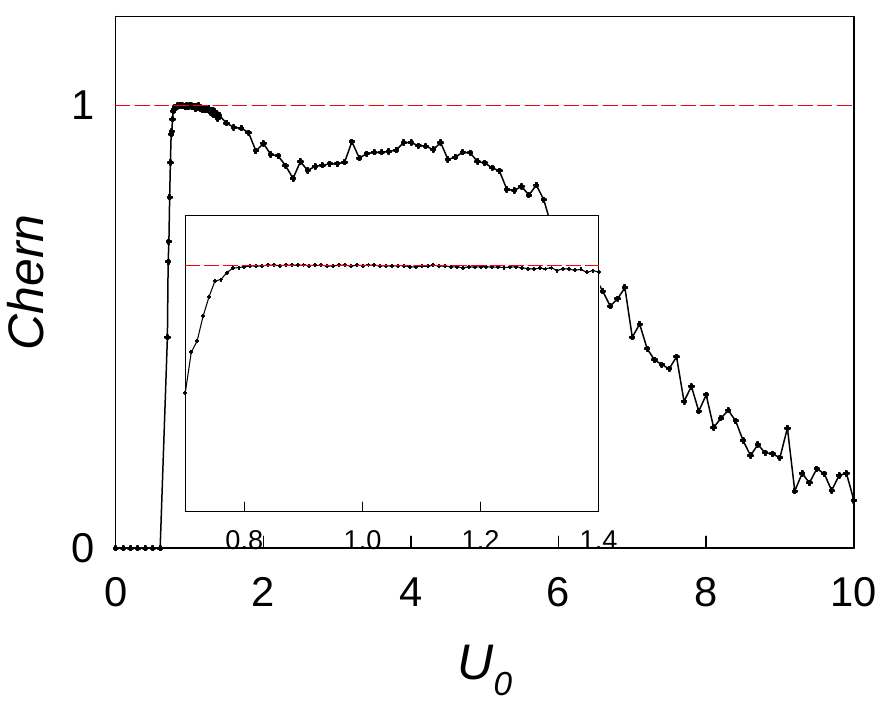}
\caption{The Chern number vs. the strength of the on-site disorder with the parameters: $B=1$, $M_0=-0.5$, $D_0=0.8$ and the length $N=100$, when the quantum pumping is trivial. The inset is the enlarged view near $U_0=1$ where the quantized plateau appears. }\label{fig6}
\end{figure}

\begin{figure}[htbp]
\centering
\includegraphics[width=9cm]{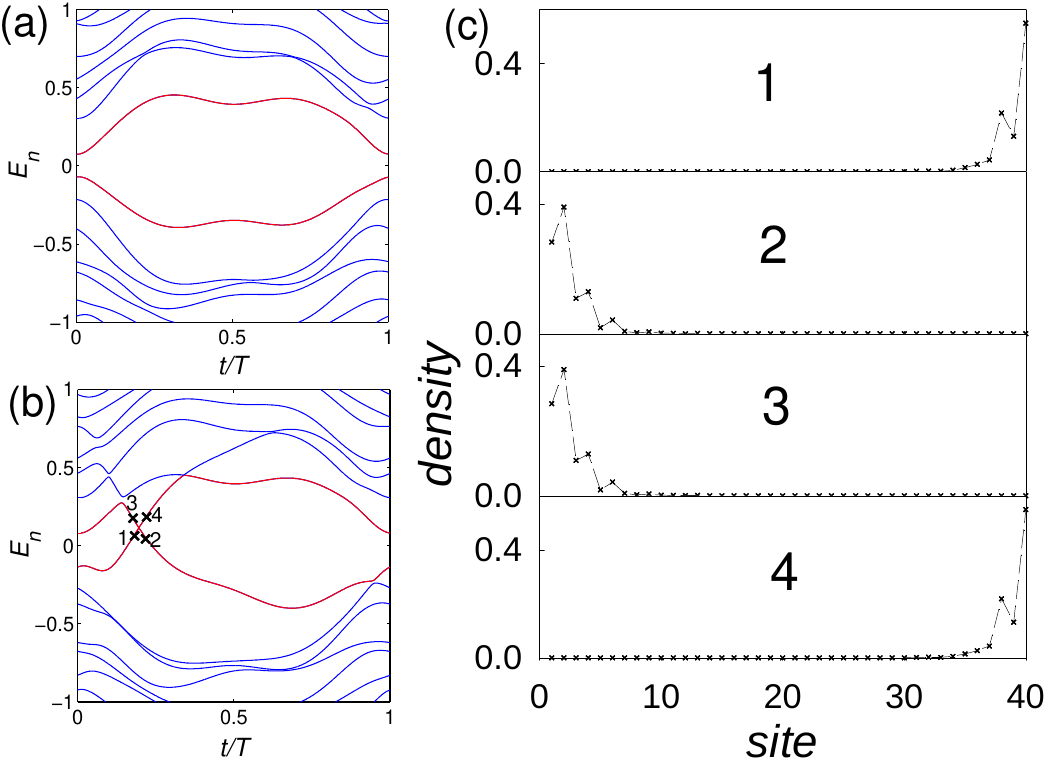}
\caption{(Color on line) The instantaneous energy spectrum: (a) on a periodic chain, (b) on an open chain. (c) the distribution of the particle on the points denoted in (b). The red lines are the $N-$ and $N+1 -$ eigenenergies. Here one random selected configuration with the disorder strength $U_0=4$ and nonzero Chern number is considered. The parameters are the same with those in Fig.\ref{fig6}, except the length $N=40$.}\label{fig7}
\end{figure}

As stated above, when the loop of the pumping process is shifted to the right of the $D-$ axis, the pumped process is trivial, i.e., no charge is pumped. It is interesting to study the effect of disorder in the trivial pumping process. Surprisingly it is found that the disorder can drive a  trivial pumping process to a nontrivial one. As an example we start from a trivial loop: $(M,D)=(M_0+0.01-M_0\cos t, D_0\sin t)$. Its crossings with the $M -$ axis are both bigger than zero, thus no in-gap states exist in the clean open chain. And also the Chern number and the pumped charge for this process are both zero. Then the on-site disorder described by Eq.(\ref{eq1}) is added. As shown in Fig.\ref{fig6}, in a finite range near $U_0=1$, the Chern numner is quantized to $1$, implying the realization of quantum pumping induced by the disorder.

When the Chern number for the cyclic process is non-zero, one manifestation is the existence of one protected crossing in the energy spectrum of the open chain. In the disordered energy spectrum  there may be many crossings. The topological protected crossing may be identified by the distribution of the particle near it. As shown in Fig.\ref{fig7}, the distribution shows the signature of the edge state near such crossing. Also for two points with similar energy and on the opposite sides of the crossing (such as $1$ and $2$, $3$ and $4$ ), the state distributes near opposite edges. Besides the instantaneous energy spectrum, we also calculate the pumped charge of the pumping process corresponding to Fig.\ref{fig7}. With properly choosing the period $T$ and the $\delta t$ in Eq.(\ref{eq4}) (to make sure that the evolution is adiabatic), the pumped charge approximates $1$. The above results are verified on different configurations with nonzero Chern number. We want to note that in the above calculations we use the disorder strength $U_0=4$ when not all configurations have nonzero Chern number, but choose randomly a configuration with nonzero Chern number. The reason for this is that near the quantized plateau $U_0=1$ the results on open chains are greatly affected by the finite-size effect due to the relatively small gap. Though $U_0=4$ is not on the plateau, the underlying physics for the configurations with nonzero Chern number induced by the disorder is the same.

\section{Conclusions}

We study a 1D topological model, which can be obtained by reducing one dimension from the BHZ model describing the 2D TI \cite{zhang}. Different from the 2D case, the 1D model uses two of the three Pauli matrix, thus the left one automatically become the protecting symmetry. By calculating the eigenenergy on an open chain and the Berry phase, we show that the 1D topological phase is stable to disorder preserving the symmetry, but fragile to disorder breaking it. We study a special case when the solution of the Hamiltonian can be easily obtained and explain the stability of the edge state to different kinds of disorders.

Then based on the 1D topological model, a stable quantum pumping is constructed. The instantaneous energy spectrum is calculated and there appears gapless states traversing the gap, which is very similar to the case in 2D topological phase. By calculating the instantaneous local current we show that an integer charge is pumped across a periodic chain in a cyclic process. Also on an open chain, an edge state can be transferred to the other edge by the quantum pumping and the process is clearly demonstrated. The Chern number can be defined for the quantum pumping and it has nonzero value for the above process. Finally we study the effect of disorder on the quantum pumping. we find that not only the quantum pumping is stable to on-site disorder, but also can be induced by it.

Experimentally adiabatic pumping has been observed, in which the tight-binding model is simulated by light propagating in the designed quasicrystal \cite{p1}.  Modulating the refraction index of the waveguides and the spacing between them can control the on-site and hopping terms. Since our model contains imaginary hopping amplitude, it is hard to realize it directly in such kind of experiment. However by performing a cyclic permutation of the Pauli
matrices $\sigma_z \rightarrow \sigma_x, \sigma_x \rightarrow \sigma_y, \sigma_y \rightarrow \sigma_z$ in Eq. (\ref{eq3}), the
Hamiltonian becomes
\begin{eqnarray}\label{eq6}
\mathcal{H''}(k,t)&=&[M+2B-2Bcos(k)]\sigma _{x} \ \\ \nonumber
&+&2Asin(k)\sigma _{y}+D \sigma_z,
\end{eqnarray}
in which the hopping amplitudes are all real. The schematic representation is shown in Fig.\ref{fig2} (c) and (d). The resulting model consists of two coupled SSH models \cite{ssh} and it may be simulated through an experimental setup composed of series of waveguides \cite{chen}. After the permutation, the main results remain. So it is very possible that these results are realized experimentally.


\section{Acknowledgements}

JQ is supported by the Fundamental Research
Funds for the Central Universities under Grant No. FRR-BR-15-009B,
the Beijing Higher Education Young Elite Teacher Project under Grant
No. 0389, and HG is supported by NSFC under Grants No.11274032 and
No. 11104189, FOK YING TUNG EDUCATION FOUNDATION and Program for NCET.


\begin{thebibliography}{99}
\bibitem{rev1}
J.E.~Moore, Nature 464 (2010) 194.

\bibitem{rev2}
M.Z.~Hasan, C.L.~Kane, \rmp 82 (2010) 3045.

\bibitem{rev3}
Xiao-Liang Qi and Shou-Cheng Zhang, \rmp 83 (2011) 1057.

\bibitem{rev4}
X.-L. Qi and S.-C. Zhang, Phys. Today 63 (2010) 33.

\bibitem{sm1}
K. Jimenez-Garcia, L. J. LeBlanc, R. A. Williams, M. C. Beeler, A. R. Perry, and I. B. Spielman, \prl 108 (2012) 225303.

\bibitem{sm2}
J. Struck, C. lschlger, M. Weinberg, P. Hauke, J. Simonet, A. Eckardt, M. Lewenstein, K. Sengstock, and P. Windpassinger, \prl 108 (2012) 225304.

\bibitem{sm3}
Y. J. Lin, R. L. Compton, K. Jim¨¦nez-Garc¨ªa, J. V. Porto, and I.B. Spielman, Nature (London) 462 (2009) 628.

\bibitem{sp1}
Y.J. Lin, K. Jimenez-Garcia, and I.B. Spielman, Nature (London) 471 (2011) 83.

\bibitem{sp2}
P. Wang, Z.-Q. Yu, Z. Fu, J. Miao, L. Huang, S. Chai, H. Zhai, and J. Zhang, \prl 109 (2012) 095301.

\bibitem{sp3}
L.W. Cheuk, A.T. Sommer, Z. Hadzibabic, T. Yefsah, W.S. Bakr, and M.W. Zwierlein, \prl 109 (2012) 095302.

\bibitem{p1}
Y.E. Kraus, Y. Lahini, Z. Ringel, M. Verbin, and O. Zilberberg, \prl 109 (2012) 106402.

\bibitem{g1}
M. C. Rechtsman, Y. Plotnik, J. M. Zeuner, D. Song, Z. Chen, A. Szameit, and M. Segev, \prl 111 (2013) 103901.

\bibitem{f1}
M.C. Rechtsman, J.M. Zeuner, Y. Plotnik, Y. Lumer, D. Podolsky, F. Dreisow, S. Nolte, M. Segev, and A. Szameit, Nature (London) 496 (2013) 196.

\bibitem{ssh}
W. P. Su, J. R. Schrieffer, and A. J. Heeger, \prl 42 (1979) 1698.

\bibitem{tp1}
Li-Jun Lang, Xiaoming Cai, and Shu Chen, \prl 108 (2012) 220401.

\bibitem{tp2}
Zhihao Xu, Shu Chen,  \prb 88 (2013) 045110.

\bibitem{tp3}
Sriram Ganeshan, Kai Sun, and S. Das Sarma, \prl 110 (2013) 180403.

\bibitem{guo1}
Huaiming Guo and Shun-Qing Shen, \prb 84 (2011) 195107.

\bibitem{guo2}
Huaiming Guo, Shun-Qing Shen, and Shiping Feng, \prb 86 (2012) 085124.

\bibitem{qp1}
Lei Wang, Matthias Troyer and Xi Dai, \prl 111 (2013) 026802.

\bibitem{exp1}
L. Fallani, J.E. Lye, V. Guarrera, C. Fort, and M. Inguscio, \prl 98 (2007) 130404.

\bibitem{exp2}
G. Roati, C.D Errico, L. Fallani, M. Fattori, C. Fort, M. Zaccanti, G. Modugno, M. Modugno, and M. Inguscio, Nature (London) 453 (2008) 895.

\bibitem{exp3}
B. Deissler, M. Zaccanti, G. Roati, C.D Errico, M. Fattori, M. Modugno, G. Modugno, and M. Inguscio, Nature Phys. 6 (2010) 354.


\bibitem{guo3}
Huaiming Guo, \pra 86 (2012) 055604.

\bibitem{bs1}
Shi-Liang Zhu, Z.-D. Wang, Y.-H. Chan, and L.-M. Duan, \prl 110 (2013) 075303.



\bibitem{bs2}
Davide Rossini, Marco Gibertini, Vittorio Giovannetti, and Rosario Fazio, \prb 87 (2013) 085131.

\bibitem{bs3}
Fabian Grusdt, Michael Honing, and Michael Fleischhauer, \prl 110 (2013) 260405.

\bibitem{mj}
Jason Alicea, Rep. Prog. Phys. 75 (2012) 076501.

\bibitem{berry}
Raffaele Resta, Rev. Mod. Phys. 66 (1994) 899.

\bibitem{qp2}
M. J. Rice and E. J. Mele, \prl 49 (1982) 1455.

\bibitem{qp3}
Liang Fu and C. L. Kane, \prb 74 (2006) 195312.

\bibitem{qp4}
Shun-Qing Shen, Topological Insulator (Springer, 2012).

\bibitem{exact}
P. Prelovsek, J. Bonca, Strongly Correlated Systems: Numerical Methods 176 (Springer, 2013).

\bibitem{note1}
The time derivative of the particle number on site $j$: $\frac{\partial c_j^{\dagger}c_j}{\partial t}=\frac{1}{i\hbar}[c_j^{\dagger}c_j,H]=\hat{{\cal J}}_j-\hat{{\cal J}}_{j+1}$, which determines the formula of the current operator.

\bibitem{ch1}
D. Xiao, M.-C. Chang, and Q. Niu, \rmp 82 (2010) 1959.

\bibitem{ch2}
T. Fukui, Y. Hatsugai, and H. Suzuki, J. Phys. Soc. Jpn. 74 (2005) 1674.

\bibitem{zhang}
B. A. Bernevig, T. L. Hughes, and S.-C. Zhang, Science 314 (2006) 1757.

\bibitem{chen}
Linhu Li, Zhihao Xu, Shu Chen,  \prb 89 (2014) 085111.
\end{thebibliography}
\end{document}